\documentclass[a4paper,fleqn]{cas-sc}

\usepackage[numbers]{natbib}
\usepackage{graphicx}
\usepackage{times}
\usepackage{booktabs}
\usepackage{threeparttable}
\usepackage{amssymb,amsmath,hyperref}
\usepackage{ulem}
\usepackage{lineno}
\usepackage{enumerate}
\usepackage{float}
\usepackage{afterpage}

\usepackage{hyperref}
\hypersetup{colorlinks=true, citecolor=blue, linkcolor=blue, urlcolor=blue}

\def\tsc#1{\csdef{#1}{\textsc{\lowercase{#1}}\xspace}}
\tsc{WGM}
\tsc{QE}

\begin{document}
\let\WriteBookmarks\relax
\def\floatpagepagefraction{1}
\def\textpagefraction{.001}

\shorttitle{Calibration and performance of DAMPE trigger system}    

\shortauthors{W.-H. Li et al.}  

\title [mode = title]{On-orbit calibration and long-term performance of the DAMPE trigger system}  



%

\author[1,2]{Wen-Hao Li}
\author[1]{Chuan Yue}
\cormark[1]
\ead{yuechuan@pmo.ac.cn}
\author[1]{Yong-Qiang Zhang}
\author[1,2]{Jian-Hua Guo}
\author[1,2]{Qiang Yuan}





\affiliation[1]{organization={Key Laboratory of Dark Matter and Space Astronomy, Purple Mountain Observatory, Chinese Academy of Sciences},
            addressline={}, 
            city={Nanjing},
            postcode={210023}, 
            state={Jiangsu Province},
            country={China}}
            
\affiliation[2]{organization={School of Astronomy and Space Science, University of Science and Technology of China},
            addressline={}, 
            city={Hefei},
            postcode={230026}, 
            state={Anhui Province},
            country={China}}





\cortext[1]{Corresponding author}



\begin{abstract}
The DArk Matter Particle Explorer (DAMPE) is a satellite-borne particle detetor for
measurements of high-energy cosmic rays and $\gamma$-rays. DAMPE has been operating smoothly in space for more than 8 years since launch on December 17, 2015. The trigger logic of DAMPE is
designed according to the deposited energy information recorded by the calorimeter. The precise calibration of the 
trigger thresholds and their long-term evolutions are very important for the scientific analysis 
of DAMPE. In this work, we develop a new method for the threshold calibration, considering the 
influence from the electronics noise, and obtain the long-term evolutions of the trigger thresholds. 
The average increase rate of the trigger thresholds for the first 4 layers of the calorimeter 
is found to be $\sim0.9\%$ per year, resulting in variations of the high-energy trigger efficiency 
of cosmic ray electrons by $\sim -5\%$ per year at 2 GeV and less than $\sim -0.05\%$ above 30 GeV. 
\end{abstract}


\maketitle

\section{Introduction}\label{Sec1}
The origin of cosmic rays (CRs) and the nature of dark matter (DM) are very important 
fundamental questions in physics and astronomy. The DArk Matter Particle Explorer 
(DAMPE; also known as ``Wukong'') is a space particle detector which is dedicated to
studies of CR physics and the indirect detection of DM via precise observations of 
high-energy CRs and $\gamma$-rays \citep{Chang2014,Chang2017}. On December 17, 2015, 
DAMPE was launched into a sun-synchronous orbit at an altitude of 500 km, and has been 
working smoothly for more than 8 years since then. The on-orbit performance of DAMPE
detector is very good and stable \citep{Ambrosi2019,Dong2019,Ma2019,Tykhonov2018,Zhang2016}, 
enabling us to achieve precise measurements of the electron plus positron spectrum 
\citep{Ambrosi2017}, proton spectrum \citep{An2019}, helium spectrum \citep{Alemanno2021a}, 
boron-to-carbon and boron-to-oxygen ratios \citep{Alemanno2022b}, and the Forbush decreases 
associated with coronal mass ejections \citep{Alemanno2021b}. DAMPE also detect a number of 
${\gamma}$-ray sources \citep{Xu2018,Duan2021,Shen2021}, and give the most stringent limits on
monochromatic ${\gamma}$-ray line emission thanks to its excellent energy resolution
\citep{Alemanno2022a,Cheng2023}. 

\begin{figure}[h]
\centering
\includegraphics[width=0.9\textwidth]{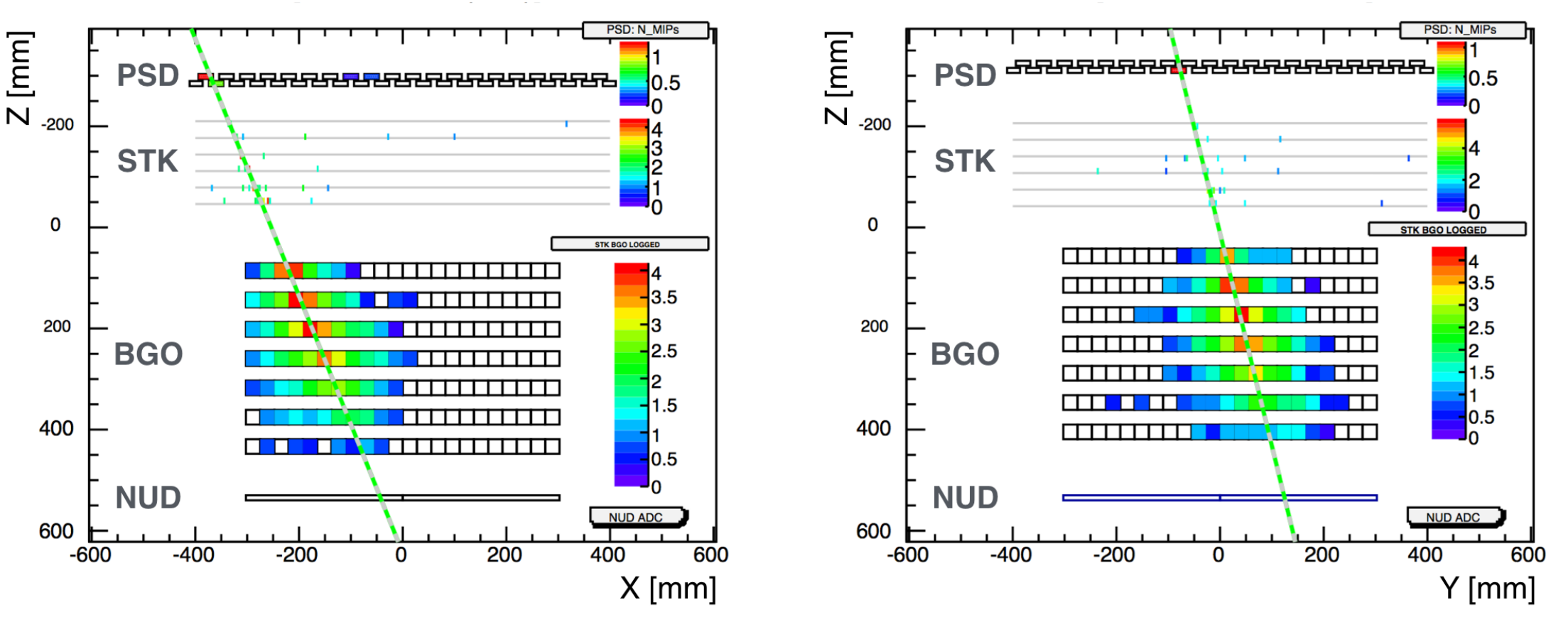}
\caption{The illustration of an electron-like event in the DAMPE detector.
The left panel is for the XZ view and the right panel is for the YZ view, with the 
Z-direction pointing from the zenith to the detector. The color bars for PSD and STK
show energy in unit of proton MIPs, and for BGO shows $\log({\rm Energy/MeV})$.
}
\label{fig1}
\end{figure}

The DAMPE payload is composed of four sub-detectors, which are, from top to bottom, 
a Plastic Scintillator strip Detector (PSD) \citep{Yu2017}, a Silicon-Tungsten 
tracKer-converter (STK) \citep{Azzarello2016}, a Bismuth-Germanium Oxide (BGO) 
imaging calorimeter \citep{Zhang2012}, and a NeUtron Detector (NUD) \citep{Huang2020}. 
A schematic view of the DAMPE payload with an example of an electromagnetic shower 
in four subdetectors are shown in Figure~\ref{fig1}. 
Combining the signals from these four sub-detectors, one can measure the charge,
direction, and energy of incident particles, and distinguish electrons/photons from
protons mainly using the shower topologies in the BGO calorimeter.

Signals from the four sub-detectors are controlled by the trigger system and synchronized 
in the data acquisition (DAQ) system \citep{Chang2017,Ambrosi2019}. The trigger system is 
designed to simultaneously achieve a high efficiency for target particles and a low global 
event rate to reduce the dead time of electronics \citep{Zhang2019}. The trigger efficiency, 
which depends strongly on the accuracy and stability of the calibrated trigger thresholds, 
is one of the most important quantities for precise flux measurements. 
Different from simply setting the threshold to the analog-to-digital converter (ADC) value corresponding to half of the 
maximum count \citep{Zhang2019}, we develop a new and more precise approach 
to determine the trigger thresholds based on the signal profile. Furthermore, we investigate 
the long-term variations of the thresholds as well as the consequence on the electron flux 
measurements in this study. 

\section{The Trigger System of DAMPE}\label{Sec2}
The trigger system of DAMPE is operated by the joint work of the BGO calorimeter and the 
trigger board. Figure~\ref{fig2} shows a brief trigger logic scheme of DAMPE. The external 
trigger is utilized in ground calibration before the launch (e.g., the beam test). 
The internal trigger is used for the on-orbit calibration of the detector. In the scientific 
operation mode, four event trigger logics, i.e., Unbiased Trigger, Minimum Ionizing Particle 
(MIP) Trigger, High Energy Trigger (HET), and Low Energy Trigger (LET), take function to 
record different types of events. A detailed trigger logic design could be found in \citep{Zhang2019}.

\afterpage{
\begin{figure}
\centering
\includegraphics[width=1\textwidth]{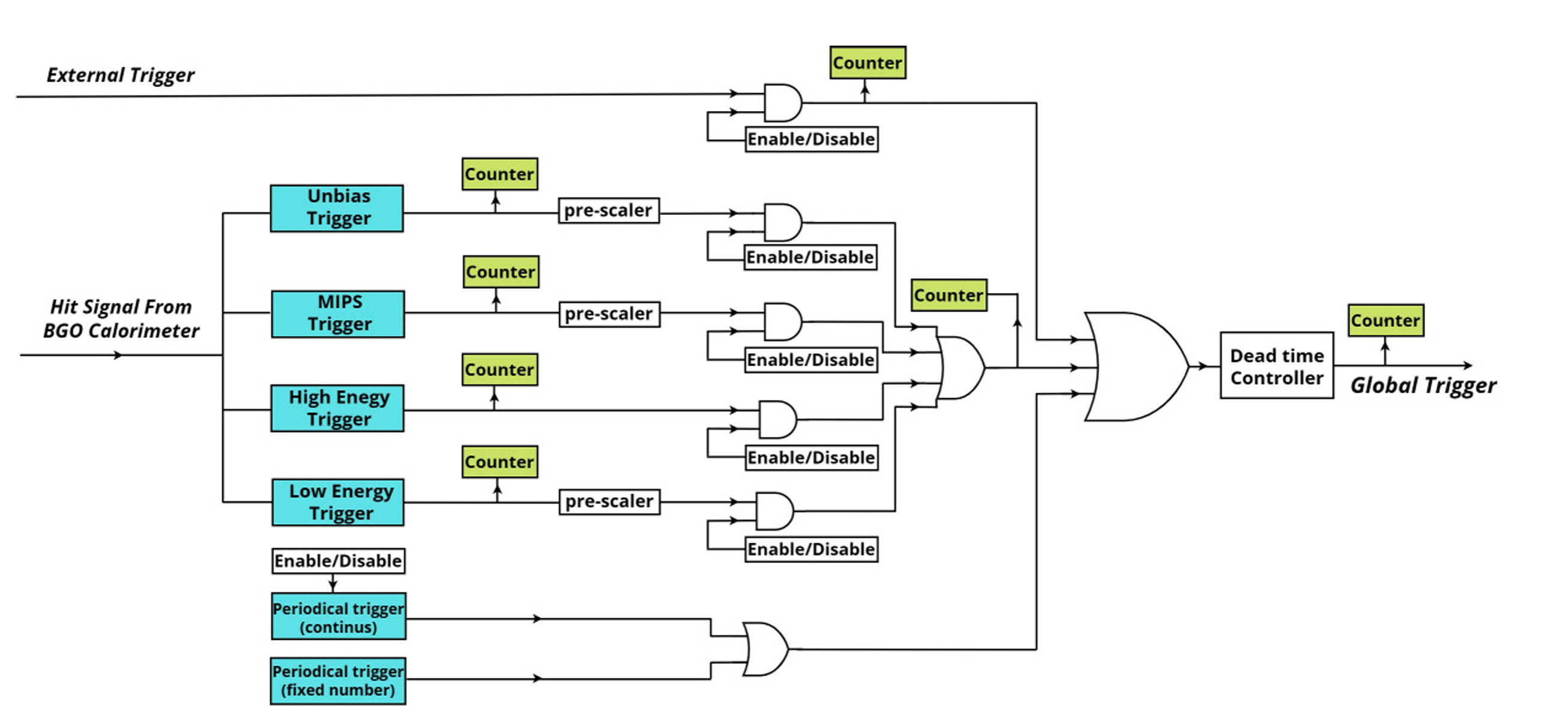}
\caption{The trigger logic of DAMPE.}
\label{fig2}
\end{figure}
}

\begin{figure}[htbp]
\centering
\includegraphics[width=0.9\textwidth]{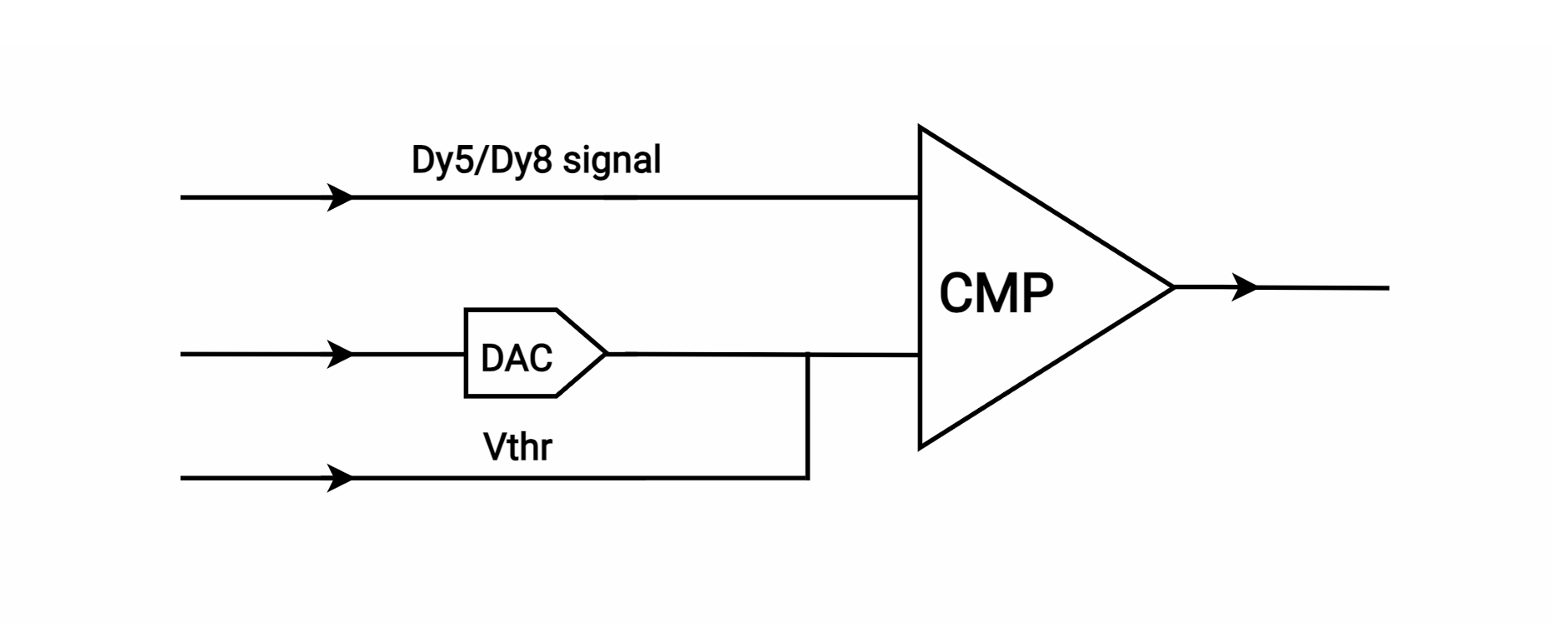}
\caption{The trigger threshold comparator (CMP) and the corresponding data flow.}
\label{fig3}
\end{figure}

\begin{figure}[htbp]
\centering
\includegraphics[width=0.9\textwidth]{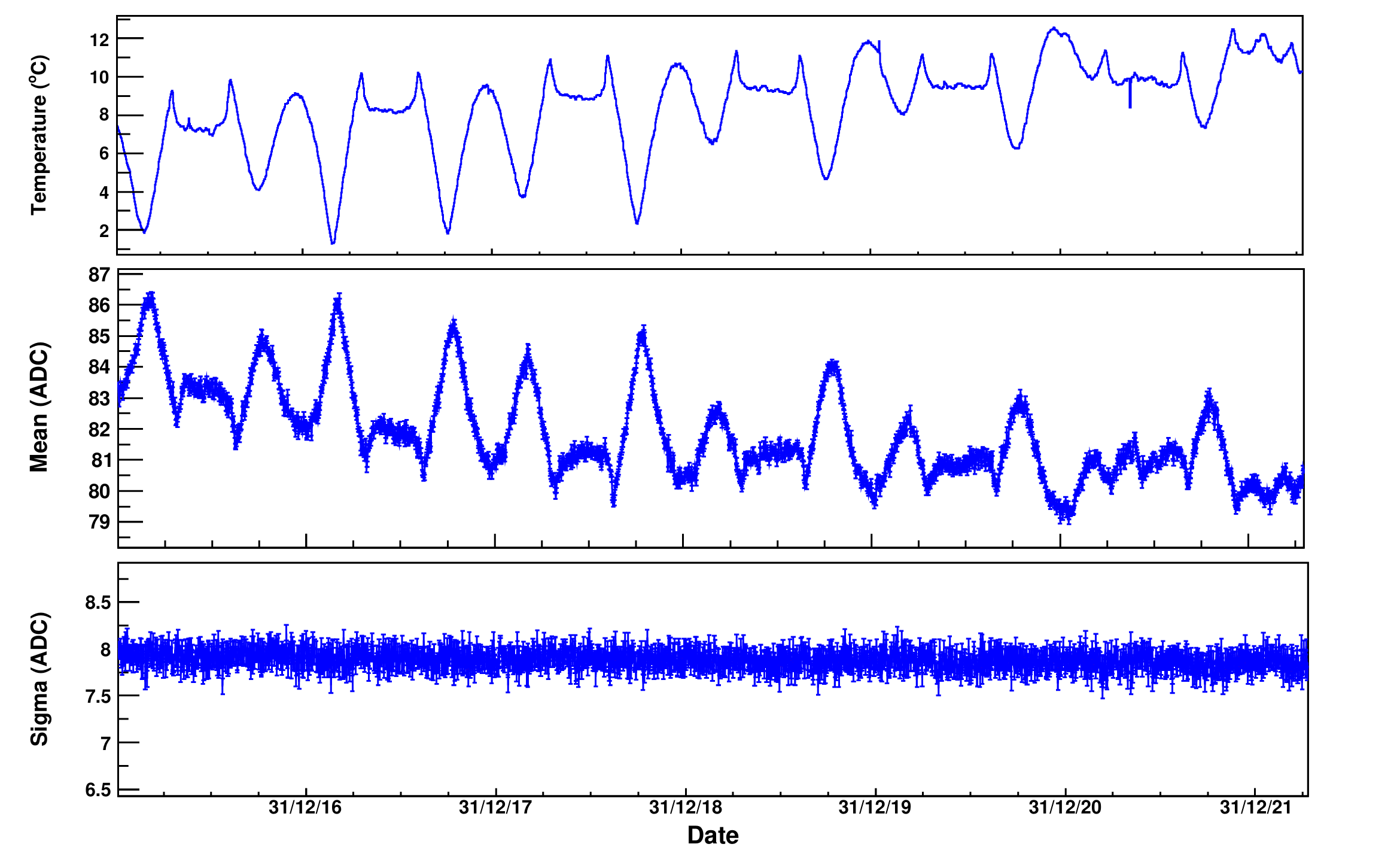}
\caption{The evolution of temperature recorded of Layer 1 Bar 3 (top panel), mean values 
(middle panel) and standard derivations (bottom panel) of pedestals for the same BGO bar over a 
long period of time, from January 2016 to June 2022.
}
\label{fig4}
\end{figure}

For different event trigger logics, different thresholds are set for different VATA chips in the front-end electronics (FEEs) of the BGO calorimeter. The final threshold is a combination of common threshold (Vthr) and a DAC threshold, as shown by Figure~\ref{fig3}.
When the electronics signal amplitude in a single BGO crystal exceeds the threshold, a hit signal will be generated 
and sent to the trigger board in the DAQ crate. If the hit signals from different layers satisfy a certain type 
of trigger logic, the trigger board will make a decision and send a trigger signal to the FEEs 
of each sub-detector to activate the digitization procedure. After that, the FEEs pack the data 
with zero-suppression and send them to the DAQ system. 
A detailed introduction on design of the readout electronics for the BGO calorimeter of DAMPE could be found in \citep{Feng2015}, and details about the DAQ system of 
DAMPE can be found in \citep{Chang2017} and \citep{Zhang2019}. 
The pedestals could be affected by many factors, and the temperature should
be one of the main factors. Figure~\ref{fig4} gives the long-term tendencies of temperature of layer 1 bar 3 (top panel), and the evolution of the mean values (middle panel) and 
widths (bottom panel) of the pedestals for the same channel of the BGO. The relatively short 
period variations of the mean values clearly reflect the temperature effect on the detector
\citep{Wang2017}. In addition, a decrease trend over years can also be seen, mainly due to 
the electronics aging and an increase in detector temperature. The standard deviations keep
relatively stable during the operation. Therefore, precisely calibrating the thresholds is 
very important in calculating the detection efficiency and the particle fluxes. 

Among the four event trigger logics, the HET which is the most widely used for the scientific 
analyses, is designed to record high energy CR or photon events with a good shower development 
in the BGO calorimeter. A set of high energy thresholds in the top four BGO layers are required 
to initiate the HET. In this work we will focus on the discussion of the HET.

\begin{figure}[htbp]
\centering
\includegraphics[width=1\textwidth]{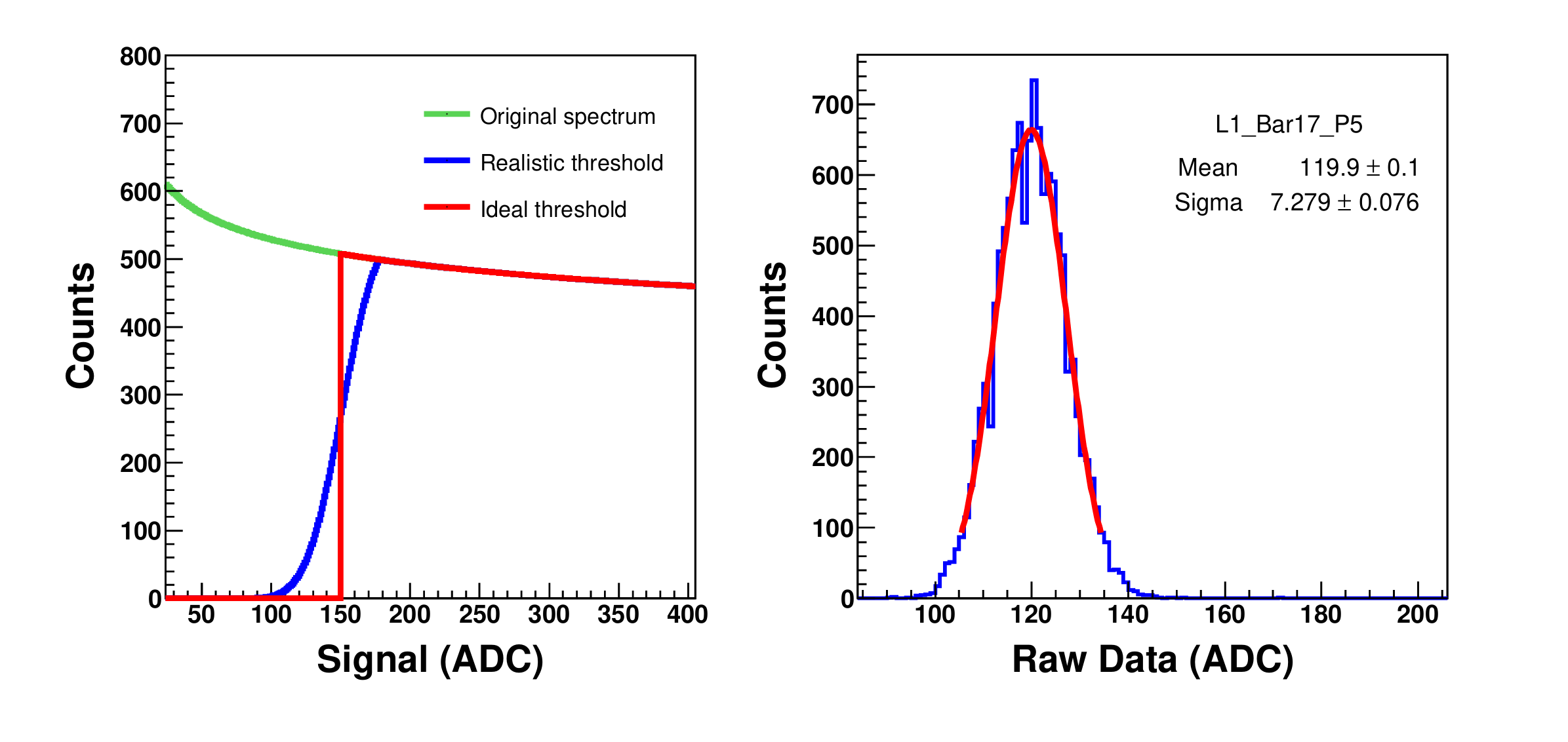}
\caption{\textbf{\textit{\textbf{L\textbf{eft}}:}} The signal ADC distribution for the original
spectrum (green line), the detected one after an ideal threshold cut (red line), and the one after
the realistic threshold cut (blue hatched).  
\textbf{\textit{Right:}}The distribution of pedestal for one particular channel (layer 1, bar 17,
P5), which can be fitted with a Gaussian function.
}
\label{fig5}
\end{figure}

\section{Threshold Calibration Method}\label{Sec3}
To cover a wide dynamic range of about $10^{6}$ for each BGO crystal, a photomultiplier tube 
(PMT) with  multi-dynode readouts is coupled to each end \citep{Zhang2012}. The scintillation 
light signal is read out from three different sensitive dynodes 2, 5, and 8 (Dy2, Dy5, and Dy8) 
of the PMT, which corresponds to low-gain, medium-gain, and high-gain channels, respectively. 
For the HET logic, specifically, the output signals from the Dy5 on the positive end (P5 channel) 
from the top three layers and the Dy8 on the negative end (N8 channel) from the fourth layer are 
employed for the hit generation \citep{Zhang2019}. 

To precisely calibrate the signal thresholds, the specific channels that used for the trigger 
decision must be firstly identified. As long as the electronical signal amplitude of specific channel in one BGO 
bar of the top four layers exceeds the corresponding threshold, this channel is the target of interest. 
Using the on-orbit data, we can obtain the readout signal distribution for such a target channel. Ideally, 
the ADC distribution should show a ``cutoff'' at the trigger threshold position, as illustrated by the
red line in the left panel of Figure~\ref{fig5}.
However, since there is electronics noise (see the bottom panel of Figure~\ref{fig4}), 
the pedestal subtracted ADC distribution shows a gradual rising shape. 
The shape is caused by the electronic noise, it influences only the trigger efficiency around the threshold, but won't change the value of the threshold.
\par
Assuming a Gaussian distribution of pedestal, the resulting signal 
distribution can thus be described as
\begin{equation}
         f(x)=f_{\rm org}(x)\int_{x_0-x}^{\infty}\frac{1}{\sqrt{2\pi}\sigma}e^{-\frac{x'^{2}}{2\sigma^{2}}}dx',
\label{fadc}
\end{equation}
where $f_{\rm org}(x)$ is the original signal distribution around the threshold cutoff, $x_0$ is
the detector threshold, $\sigma$ is the standard deviation of the pedestal. The integral of the Gaussian 
function represents the probability that the fluctuation of pedestal can compensate the signal and make
the total readout exceed the threshold. The above formula can reduce to $f(x)=f_{\rm org}(x)H(x-x_0)$ 
for the ideal case when $\sigma \to 0$, where $H(x-x_0)$ is the step function.

\section{Results and Discussion}\label{Sec4}
\subsection{Trigger thresholds}
We identify the specific channels that participate in the trigger decision for each HET event. 
After that, we get the ADC distribution of each channel using events accumulated in one day. 
Then the trigger threshold of each channel is obtained by fitting the readout signal distribution with 
Eq.~(\ref{fadc}), assuming a power-law form of $f_{\rm org}(x)$. 
In a wide energy range, the spectrum of cosmic rays is not a simple power law due to the geomagnetic field. However, in a relatively narrow range around the threshold, a power law function may well approximate the spectral behavior. As an assessment of the possible effect due to the assumed original spectral shape, we test the other form of a log-parabola spectrum, and the differences between power-law and log-parabola are very small.
\par
\begin{figure}[htbp]
\centering
\includegraphics[width=0.9\textwidth]{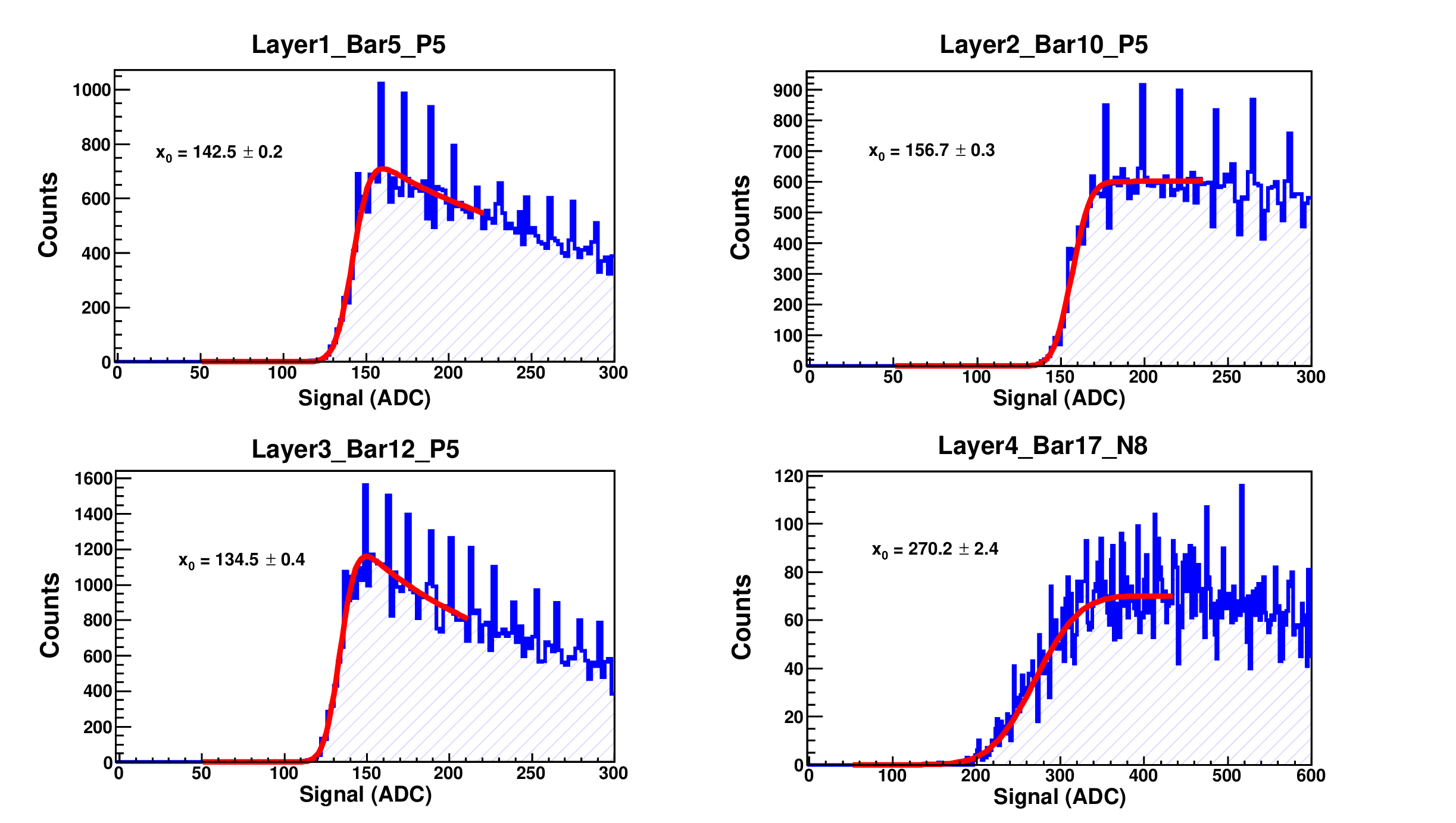}
\caption{The ADC Signal distributions (histogram) and fitting results (red lines) for selected channels of the first four BGO layers. Layer1${\_}$Bar5${\_}$P5 refers to the positive dynode 5 of the 5th bar in the fisrst layer. N is the normalization parameter, ${x_{0}}$ is the ADC threshold given by the fit.}
\label{fig6}
\end{figure}

The ADC peak position for each channel is given by a adaptively pre-search and the fit upper limit is a little higher than the peak position as we assume the ADC distribution shows a power-law in a reasonable signal range.
The ADC distributions of on-orbit data and the fitting results for four channels are shown in
Figure~\ref{fig6}. The rising profile of the signal distribution around the threshold can be well fitted by the theoretical function, Eq.~(\ref{fadc}).
It should be noted that the power law variables are all set as free parameters in the fit of the daily data considering the solar modulation on the cosmic ray spectrum.
The uncertainty of the fitted threshold value is typically $\sim$ 0.2 ADC, which is 5 times smaller than the uncertainty of $\sim$ 1 ADC from 
the direct counting method.

\begin{figure}[htbp]
\centering{
\includegraphics[width=0.8\textwidth]{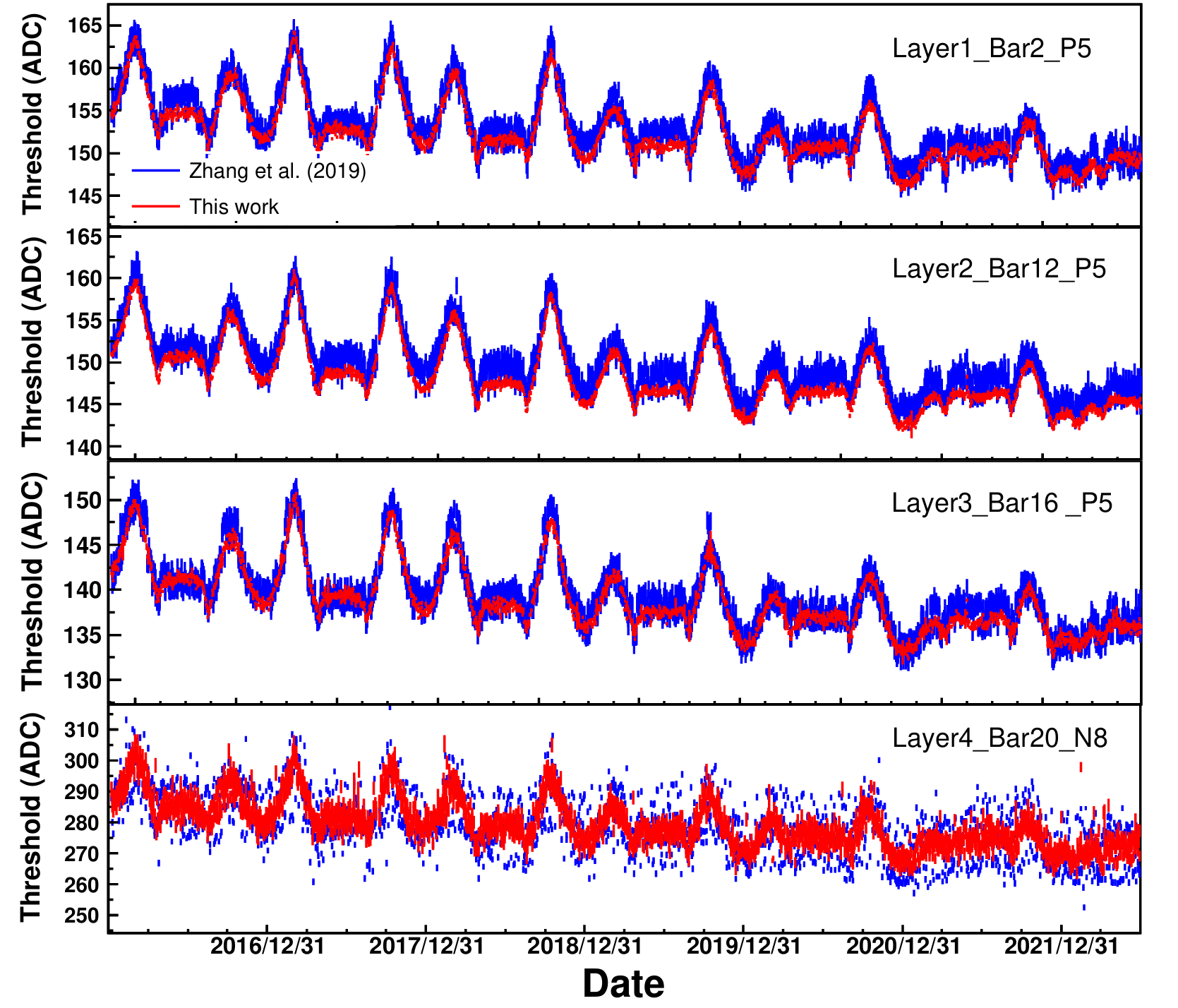}}
\caption{Time evolution of the trigger thresholds from January 2016 to June 2022. 
One channel is shown in each layer from layer 1 to layer 4 from top to the bottom histogram. Red dots show the results obtained in this work, 
and blue ones are results from the old method \citep{Zhang2019}.}
\label{fig7}
\end{figure}

One can note from Figure~\ref{fig5} that, there are some discrete spikes in the ADC distributions,
which are from the differential nonlinearity of the AD976 chip \citep{AD976} implemented in the 
electronics readout of the PMT. These spikes differ from one to another. Statistically they can be
averaged when calculating the total energy deposit for an event, thereby such an effect does not 
result in significant bias on the energy measurement. However, if the spikes are located in the 
rising part of the distribution, they will strongly affect the threshold determined with the 
traditional method, i.e., the ADC value at half of the maximum. On the other hand, the fitting 
method is less sensitive to those spikes, and the obtained threshold value is more robust.

The time evolution of the trigger thresholds for one channel in each of the first four years 
are given in Figure~\ref{fig7}, from January 2016 to June 2022. The results derived with the traditional method
\citep{Zhang2019} are also shown for comparison. The traditional method regard the spectrum before the threshold as a constant value, which is unreasonable. Therefore we use a power-law spectrum representing the cosmic ray spectrum around the threshold within a reasonable range.
Our new method gives more accurate and smaller scatterings of the
thresholds especially, compared with the old method. Besides the relatively short-term variations which mainly
correlate with temperature \citep{Wang2017}, long-term declines of the thresholds for all the four
layers are visible. 

\begin{figure}[htbp]
\centering
\includegraphics[width=0.9\textwidth]{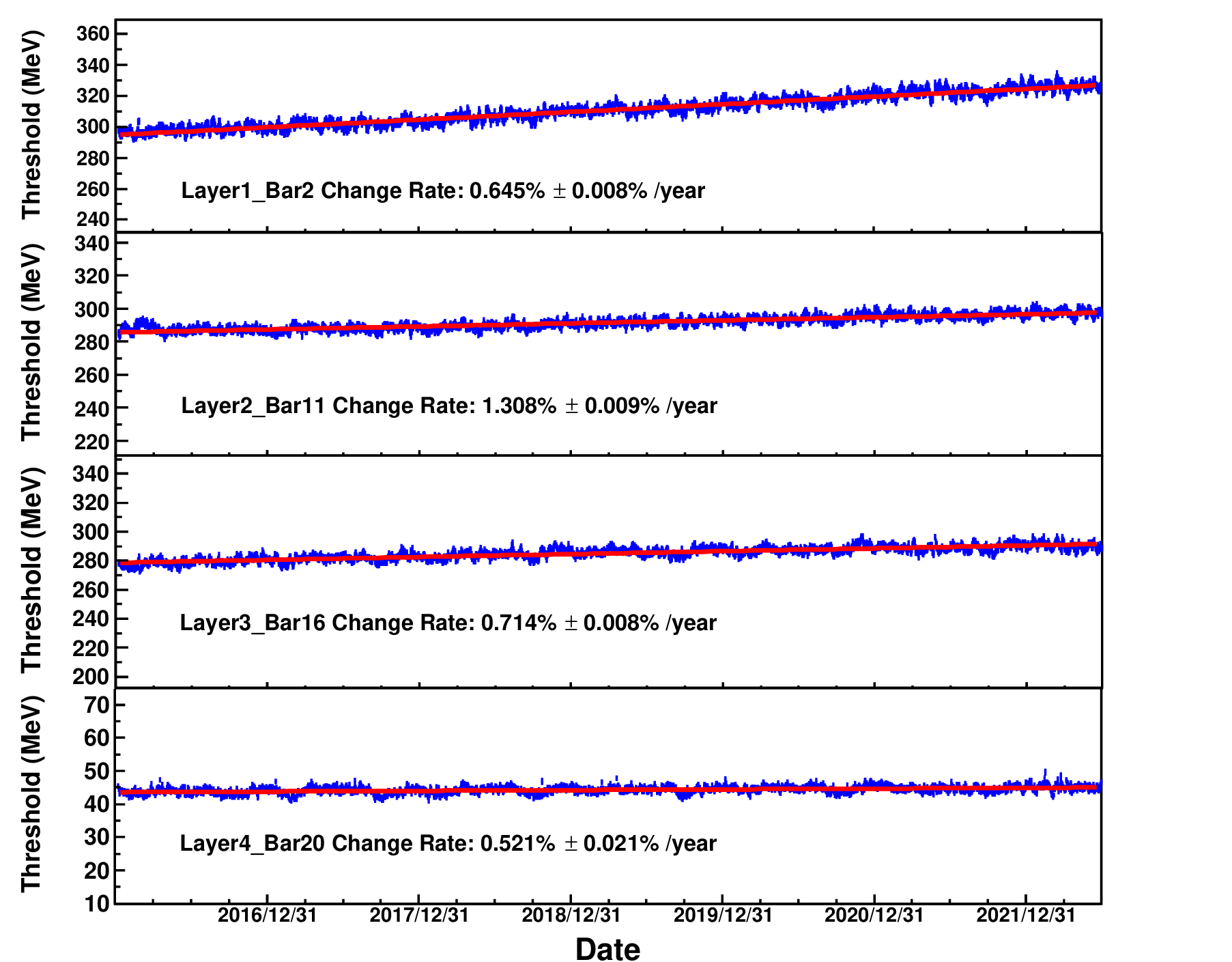}
\caption{Time evolution of the trigger thresholds in energy, from January 2016 to June 2022 for four channels 
in the first four BGO layers.}
\label{fig8}
\end{figure}

\begin{figure}
\centering
\includegraphics[width=0.8\textwidth]{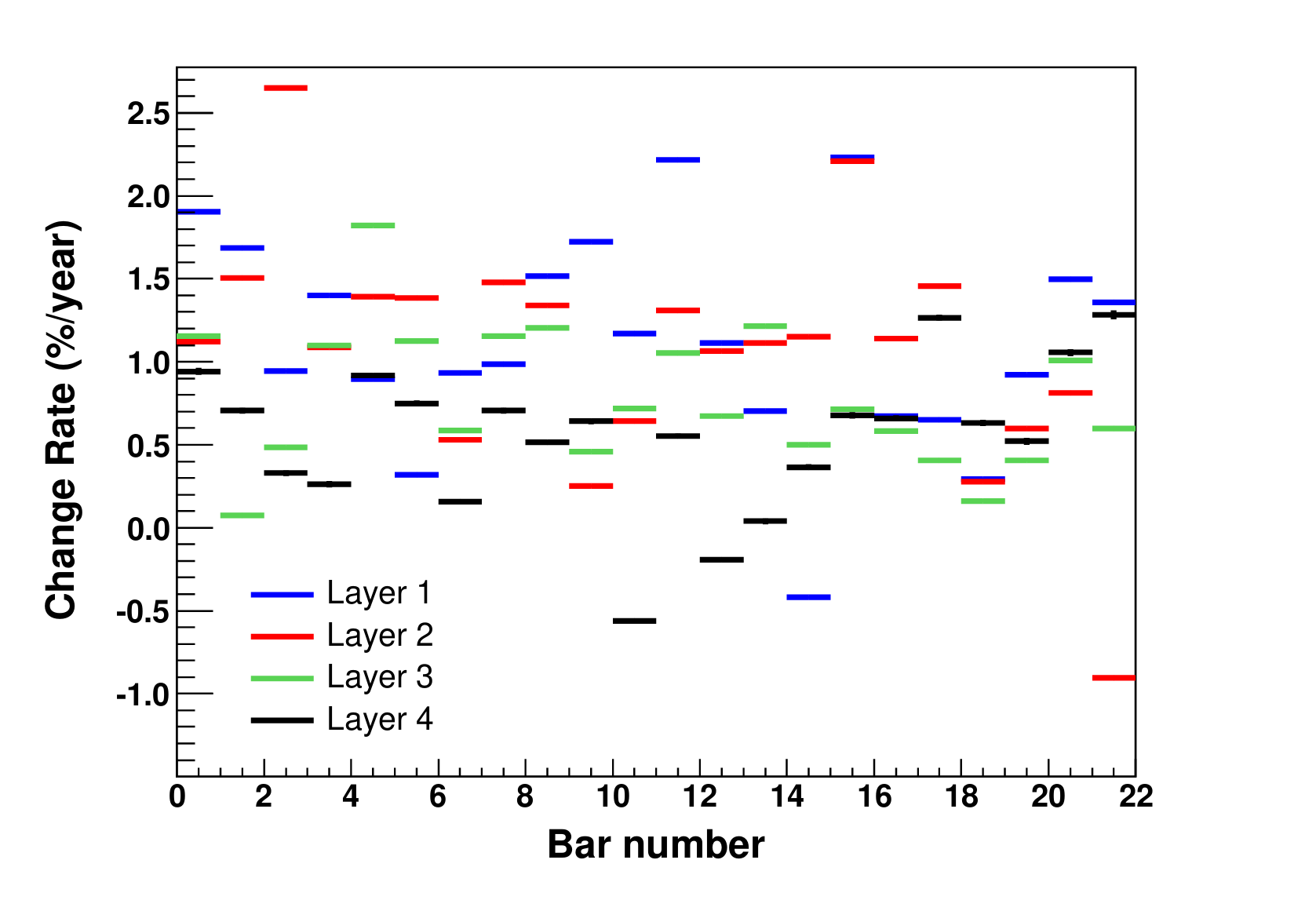}
\caption{The change rates of the trigger thresholds in energy for different BGO bars.}
\label{fig9}
\end{figure}

To evaluate the long-term variation more precisely, we convert the trigger thresholds in ADC unit 
to deposited energy (MeV), considering the gain of PMTs and removing the temperatue effect\footnote{There
is a strong correlation between the scintillation light yield and temperature, which results in 
variations of the ADC values of the proton MIP spectrum. Through re-scaling the peak ADC value to
the expected peak energy, approximately 23 MeV, we can effectively eliminate the temperature effect.}.
In daily on-orbit calibrations, the PMT gain is characterized by the peak of the proton MIP 
spectrum of each Dy8 channel \citep{Wang2017}. The trigger threshold energy is thereby calculated from Dy8 as

\begin{equation}
E_{\rm tri}= {\rm ADC}_{\rm tri} \cdot \frac{E_{\rm MIP}}{{\rm ADC}_{\rm MIP}}, 
\label{Eq2}
\end{equation}

For the Dy5 channel, the ADC value needs to be 
converted to the Dy8 ADC value with the dynode-58 conversion parameters, the trigger threshold is thereby calculated as:

\begin{equation}
E_{\rm tri}= ({\rm ADC}_{\rm tri} \cdot k_{58}+b_{58}) \cdot \frac{E_{\rm MIP}}{{\rm ADC}_{\rm MIP}}, 
\label{Eq3}
\end{equation}

where $k_{58}$ and $b_{58}$ are the slope and the intercept of the dynode-58 relation, 
${\rm ADC}_{\rm MIP}$ is the peak ADC value of the on-orbit proton MIP spectrum, and $E_{\rm MIP}$ 
is the peak energy of the simulated proton MIP spectrum. 
\par

\par

The long-term evolutions of the trigger threshold energy are shown in Figure~\ref{fig8}. We can find that the threshold ADC of the first three layers is half of the fourth layer, while the threshold energy is about five times that of the fourth layer.

This is due to that the desired threshold is 10 MIPs for the first three layers and 2 MIPs for the fourth layer, therefore the thresholds in MeV unit differ by a factor of 5. However, the difference of threshold in ADC value is a combined result from the trigger logics, as well as the photo-attenuation sheet placed in different sides. 
The high energy trigger logic set is (L1 P5 \& L2 P5 \& L3 P5 \& L4 N8), the gain of Dy8 is about 30 times larger than that of Dy5. The attenuation coefficient of the photo-attenuation sheet attached in the positive side is about 5 times smaller than the negative side. All these effects lead to the differences of the thresholds signals between the first 3 layers and the 4th layer.
\par
The threshold energy increases with time as shown in Figure~\ref{fig8}, due to particles with the same energy need a higher pedestal up-fluctuation to exceed the threshold, hence the trigger
efficiency becomes lower. 
Using a straight line to fit the long-term evolution, we obtain the relative variation rate of 
each channel for each bar of the first four layers, as shown in Figure ~\ref{fig9}. For most of 
channels, the trigger threshold shows a positive increase rate. The average increase rates for 
different layers are given in Table 1. We can see that for deeper layers the average increase rate seems smaller, however, the change rate of different bars in the same layer differs very much, causes a relative large error on the average increase rate, so the difference between the first three layers are not very significant. 
The overall increase rate of trigger thresholds of energy is $\sim0.9\%$/year, resulting in a decrease of 
the on-orbit trigger efficiency of DAMPE. 

Different from the results threshold ADC shown in Figure~\ref{fig7}, the threshold energy increases 
with time.
The general increase of the trigger thresholds is primarily due to a combined influence from the aging of the BGO crystals and the related electronics. The relevant reasons could be concluded as below:
(1) The radiation damage of the BGO crystals, causing a decrease of the scintillation light production; (2) The decay of the PMT gain; (3) The long-term changing of the electronics.
In addition, as the on-orbit temperature of BGO shows a increase trend, which could induce a long-term change on the scintillation light production of the BGO crystals, thereby   affecting the trigger threshold. These above effects are mixed together to result in the long-term variations of the trigger thresholds. So it is very difficult to preciously calculate the influences from each reason mentioned above.

\begin{figure}[htbp]
\centering
\includegraphics[width=1\textwidth]{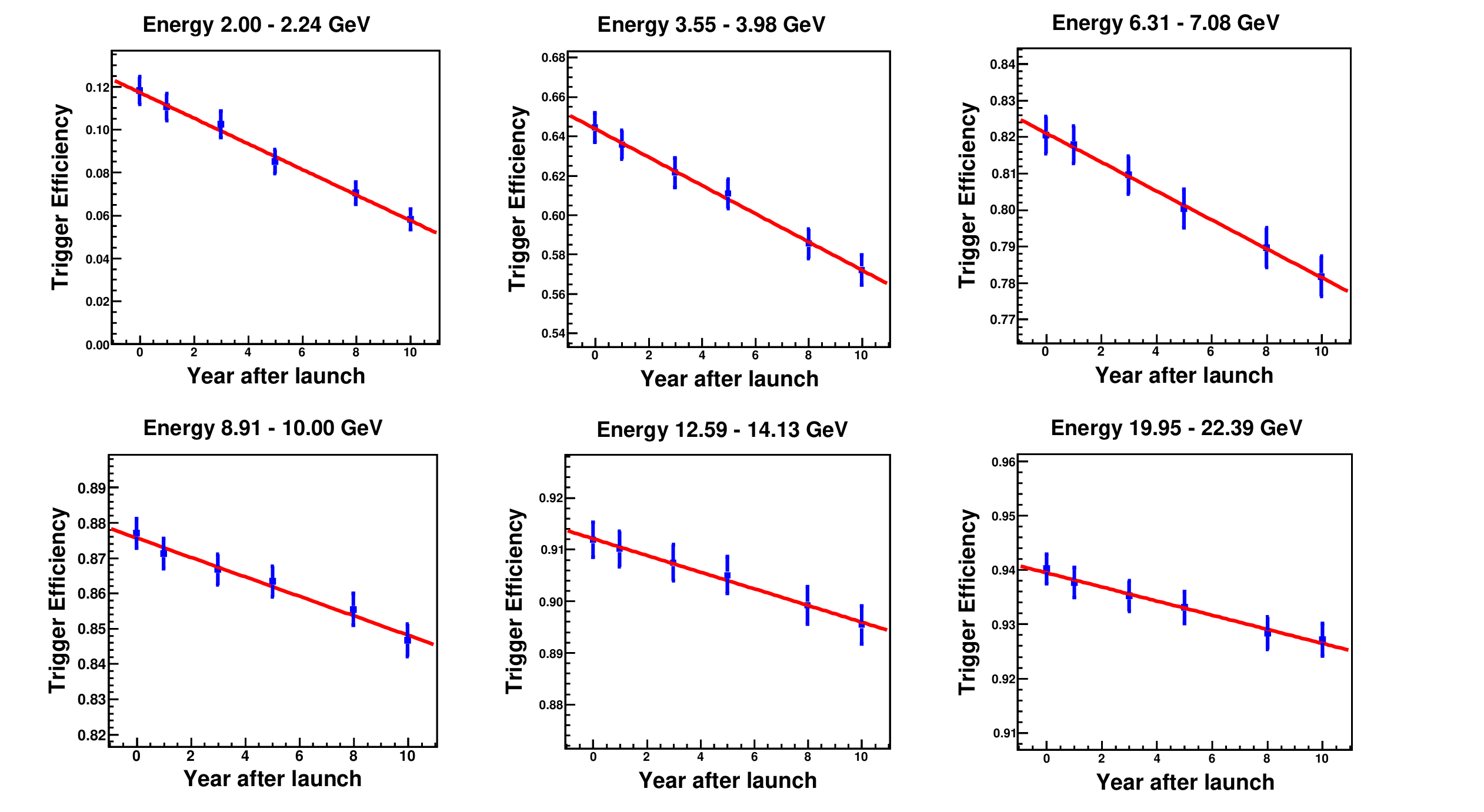}
\caption{
Time evolution of simulated electron trigger efficiencies in different energy bins, data based on Geant4 simulation.}

\label{fig10}
\end{figure}

\begin{table}[htbp]
\centering
\begin{tabular}{ccccc}
\hline\hline
    & Layer 1  & Layer 2  & Layer 3  & Layer 4 \\ \hline
Daily increase rate (day$^{-1}$) & $3.07\times10^{-5}$ & $2.93\times10^{-5}$ & $2.14\times10^{-5}$ & $1.52\times10^{-5}$ \\ \hline
Yearly increase rate (year$^{-1}$) & 1.10\% & 1.09\% & 0.73\% & 0.65\% \\ \hline\hline
\end{tabular}
\label{table1}
\caption{Average increase rates of the trigger thresholds in energy for the first four BGO layers.}
\end{table}

\subsection{Impact on electron trigger efficiency}
Taking the HET efficiency for electrons as an example, we study the impact of variations of 
the trigger thresholds on the efficiency calculation using the Monte Carlo (MC) simulation data. The MC simulation is performed with the GEANT4 tookit \citep{Allison2016} based on an accurate geometric model including both the payload and the satellite platform. 
In general, the simulation process includes particle generation, transportation simulation, digitization and reconstruction \citep{Jiang2020}. In the digitization, we add the electronic responses and convert the the raw hits in each detection unit into digital signals, which have same format as the raw flight data. The real trigger logics with the energy thresholds (MeV) as obtained above are then applied to the digitized data. By applying different trigger thresholds at specific time in the MC data, the long-term variation of the trigger efficiency can be estimated.
\par
\begin{figure}
\centering
\includegraphics[width=0.9\textwidth]{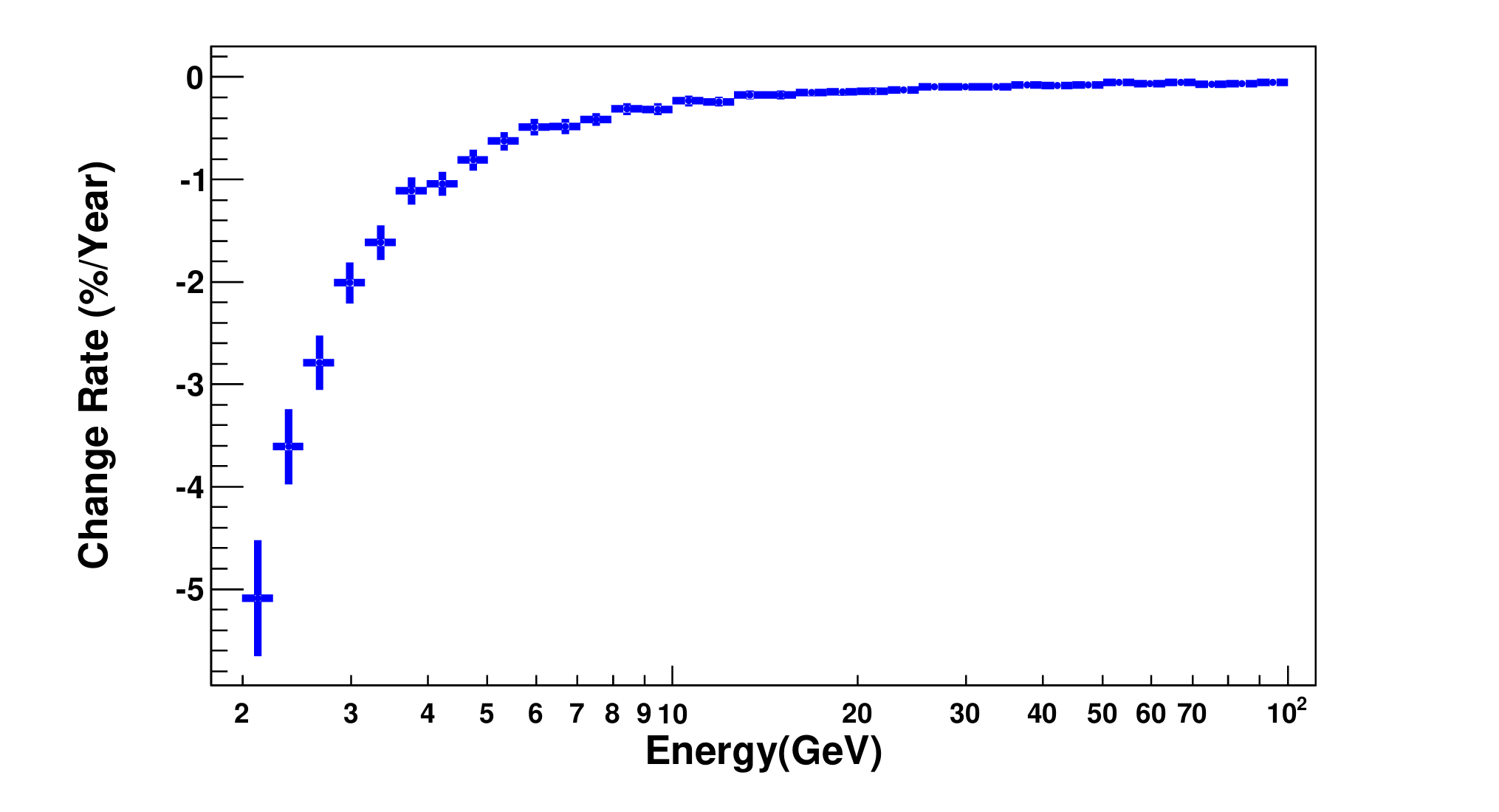}
\caption{Energy dependence of the variation rate of the HET efficiency for electrons.}
\label{fig11}
\end{figure}

The time dependence of the HET efficiency for MC electrons in different energy bins 
are shown in Figure~\ref{fig10}. For all energy bins, the HET efficiency shows clear decrease 
trends, as can be expected from Figure~\ref{fig8} that the threshold energy increases with time. 
The change of the trigger efficiency can be well described by a linear function (see the red solid
lines in Figure~\ref{fig8}). The energy dependence of the variation rate of the HET efficiency 
is shown in Figure~\ref{fig11}. The variation rate is $\sim-5\%$/year at 2 GeV, $-0.15\%$/year 
at 20 GeV, and decreases to $-0.05\%$ above 30 GeV. 
The CR electron spectra measurement above 20 GeV is thus not significantly affected by the time 
variations of the trigger thresholds. However, for the low-energy measurements, such as the time variations of cosmic-ray electron fluxes associated with solar activity \citep{Alemanno2021b} and the long-term analyses of GeV ${\gamma}$-ray sources, 
the evolution of the trigger efficiency must be properly addressed.

\section{Conclusion}\label{Sec5}
The on-orbit trigger system of DAMPE has been working stably since the launch. The trigger threshold 
calibration is a key issue for the precise calculation of the trigger efficiency. In this work, we
develop an adaptive fitting method for the trigger threshold calibration via considering the influence 
from the electronics noise. Compared with the traditional method which simply adopts the ADC value
at half of the maximum number of counts, this new method is more precise and less sensitive to the
differential nonlinearity of the chips. 
\par
We obtain the trigger thresholds of HET and their long-term
evolution. The results shows that the threshold energy mostly show an increase trend with time, 
with an average increase rate of $\sim0.9\%$/year. The change rate for the four layers utilized in the HET indicates that such long-term evolution 
could be caused by many reasons such as changes in related electronics and radiation damage in space. As a result, the HET efficiency of electrons shows a decrease trend. The time variation rate of the HET efficiency
is $\sim-5\%$/year at 2 GeV and $-0.15\%$/year at 20 GeV.
\par
The variation of the high trigger efficiency needs to be taken into account when it comes to research such as ${\gamma}$-ray  analysis, observations of fine time structures in the cosmic ray fluxes, when the high energy trigger is applied.

\section*{Declaration of competing interest}
The authors declare that they have no known competing financial interests or personal relationships that could have appeared to
influence the work reported in this paper

\section*{Data availability}
Data will be made available on request.

\section*{Acknowledgements}
This work is supported by the National Key Research and Development Program of China (No. 
2022YFF0503301
), the National Natural Science Foundation of China (Nos. 
12220101003, 12173099, 12003075, 12227805), the CAS Project for Young Scientists in Basic Research (No. 
YSBR-061, YSBR-092), the Strategic Priority Program on Space Science of Chinese Academy of Sciences (No. 
E02212A02S), the Youth Innovation Promotion Association CAS, the Young Elite Scientists Sponsorship Program by CAST (No. YESS20220197).


\printcredits

\bibliographystyle{cas-model2-names}




\end{document}